\documentclass[aps,prc,preprint,amsmath,showpacs,english]{revtex4}

\usepackage[T1]{fontenc}
\usepackage[latin1]{inputenc}
\usepackage{float}
\usepackage[dvips]{color,graphicx}
\usepackage{babel}
\makeatother

\begin{document}

\title{Quark-Hadron Duality and Truncated Moments of Nucleon Structure
	Functions}

\author{A. Psaker$^{1,2,3}$, W. Melnitchouk$^1$, M.~E.~Christy$^2$,
	C.~Keppel$^{1,2}$}
\affiliation{
	$^1$ Jefferson Lab, 
	     Newport News, Virginia 23606, USA			\\
	$^2$ Hampton University, Hampton, Virginia 23668, USA   \\
	$^3$ American University of Nigeria, Yola, Nigeria}

\begin{abstract}
We employ a novel new approach to study local quark-hadron duality
using ``truncated'' moments, or integrals of structure functions over
restricted regions of $x$, to determine the degree to which individual
resonance regions are dominated by leading twist.
Because truncated moments obey the same $Q^2$ evolution equations as
the leading twist parton distributions, this approach makes possible
for the first time a description of resonance region data and the
phenomenon of quark-hadron duality directly from QCD.
\end{abstract}

        
\maketitle

\section{Introduction}
\label{sec:intro}

The structure and interactions of hadrons at intermediate energies 
represents one of the most outstanding problems in the standard model
of particle and nuclear physics.
Many hadronic observables can be described at low energies in terms of 
effective, hadronic (meson \& baryon) degrees of freedom, while at high 
energies perturbative QCD has proved a highly successful approach to 
describing phenomena in terms of elementary quark and gluon constituents.
The precise nature of the transition between the two regimes has remained 
shrouded in mystery, however, and represents a fundamental challenge to 
our understanding of strong nuclear interactions within QCD.

One of the most intriguing connections between the low and high energy 
realms is the phenomenon of quark-hadron duality, in which one finds in 
certain cases dual descriptions of observables in terms of either explicit
quark degrees of freedom, or as averages over hadronic variables.
A spectacular example of such a duality is in inclusive
electron--nucleon scattering.
First observed by Bloom \& Gilman in the early days of deep inelastic
scattering (DIS) measurements \cite{BG}, this duality manifests itself
in the similarity of structure functions averaged over the resonance
region (which is characterized by hadronic bound states) and the
scaling or leading twist function describing the high energy, deep
inelastic continuum (characterized by scattering from free quarks)
\cite{MEK}.

Unraveling the origin of the ``Bloom-Gilman'' duality from first
principles has proved to be a major challenge in QCD.
Until now the only rigorous connection with the fundamental theory
has been within the operator product or ``twist'' expansion,
in which moments of structure functions are expanded as a series
in inverse powers of the virtuality $Q^2$ of the exchanged photon.
The leading, ${\cal O}(1)$ term is given by matrix elements of
(leading twist) quark-gluon bilocal operators, and is associated with
free quark scattering, while the ${\cal O}(1/Q^2)$ and higher terms
correspond to nonperturbative (higher twist) quark-gluon interactions.
Bloom-Gilman duality is then interpreted in this language as the
suppression of higher twist contributions to the moments \cite{DGP}.

Recent experimental data \cite{Niculescu} suggest that not only moments,
but structure functions in individual resonance regions, such as the 
$\Delta$, $S_{11}$ or $F_{15}$ regions, closely resemble the leading 
twist structure functions over the same intervals.
This indicates that duality also exists locally, in restricted regions
of hadronic final state mass $W$.
The appearance of this {\em local} duality cannot, however, be explained
with the theoretical tools presently at our disposal, and insight into 
the workings of duality for individual resonance regions has been 
confined to QCD-inspired models of the nucleon.
As such our understanding of quark-hadron duality in nucleon structure 
functions within QCD is incomplete.

In this paper we present a new approach to the study of local 
quark-hadron duality within a perturbative QCD context, using 
``truncated'' moments of structure functions.
The virtue of truncated moments is that they obey a similar set of
$Q^2$ evolution equations as those for parton distributions 
\cite{Forte,Kotlorz}, which therefore enables a rigorous connection
to be made between local duality and QCD.
It allows us to quantify for the first time the higher twist content
of various resonance regions, and determine the degree to which 
individual resonances are dominated by leading twist.

Truncated moments were introduced several years ago by Forte {\em et al.}
\cite{Forte} to study structure function moments for which small-$x$ 
data were not available.
By restricting or truncating the integration region to some minimum
value of the Bjorken $x$ variable, one could avoid the problem of 
extrapolating parton distributions into unmeasured regions at small $x$.
Later Kotlorz \& Kotlorz \cite{Kotlorz} developed an alternative
formulation of the evolution equations which avoids the problem of
mixing of higher truncated moments when evolving in $Q^2$.

In this work we partially follow the latter approach and apply it to the
study of structure functions in the {\em large-$x$} region, populated by 
nucleon resonances.
In particular, we study the $Q^2$ evolution of structure functions
integrated over specific nucleon resonance regions.
To facilitate such an analysis requires extension of the definition
of the truncated moments to include both upper and lower truncations.
We show that these ``doubly truncated'' moments also obey the same
$Q^2$ evolution equations.
Using recent high-precision data on the proton $F_2$ structure function 
from Jefferson Lab and elsewhere, we quantify the size of the higher
twists for the lowest three moments in various regions of $W$.
This represents the first quantitative test of local duality in
structure functions within a QCD framework.

This paper is organized as follows.
In Sec.~\ref{sec:evolve} we review the essential elements of $Q^2$ 
evolution via the DGLAP equations, and introduce truncated moments
together with their evolution.
We test the accuracy of our numerical evolution procedure in
Sec.~\ref{sec:data}, and further study recent proton structure function
data in the nucleon resonance region at $W < 2$~GeV. 
We divide the data into the three traditional resonance regions and 
extract the leading and higher twist content of each region.
Finally, in Sec.~\ref{sec:conclusion} we summarize the conclusions
of this analysis and outline future work.

\section{Truncated Moments and Evolution}
\label{sec:evolve}

\subsection{QCD Evolution Equations}

The $Q^2$ dependence of a parton distribution function (PDF) $q(x,Q^2)$
is described in perturbative QCD (pQCD) by the DGLAP evolution equations
\cite{DGLAP}:
\begin{eqnarray}
\frac{dq\left(x,Q^{2}\right)}{dt} & = & 
\frac{\alpha_{S}\left(Q^{2}\right)}{2\pi}
\left(P\otimes q\right)\left(x,Q^{2}\right)\ ,
\label{eq:evolutionequation}
\end{eqnarray}
where $t \equiv \ln\left(Q^2/\Lambda_{\rm QCD}^2\right)$,
with $\Lambda_{\rm QCD}$ the QCD scale parameter,
and the symbol $\otimes$ denotes the Mellin convolution,
\begin{eqnarray}
\left(P\otimes q\right)\left(x,Q^{2}\right) & = & 
\int_{x}^{1}\frac{dy}{y}\;
P\left(\frac{x}{y},\alpha_{S}\left(Q^{2}\right)\right)
q\left(y,Q^{2}\right)\ ,
\label{eq:convolution}
\end{eqnarray}
between the parton distribution $q$ and the splitting function
(or the evolution kernel) $P$.
In pQCD the latter can be expanded as a series in the strong running
coupling constant $\alpha_{S}\left(Q^{2}\right)$.
For the nonsinglet (NS) case, $q$ is one of the flavor nonsinglet
combinations of quark distributions and $P$ the corresponding NS
splitting function. 
For the singlet case, on the other hand, $q$ is a vector whose 
components are the flavor singlet combination of quark distributions 
and the gluon distribution, and correspondingly $P$ is a $2\times2$
matrix of splitting functions.

Taking moments, the convolution in Eq.~(\ref{eq:evolutionequation})
turns into an ordinary product, and the evolution equations become
ordinary first order differential equations in moment space $n$,
\begin{eqnarray}
\frac{d{\cal M}_n(Q^2)}{dt} & = & 
\frac{\alpha_S(Q^2)}{2\pi}\;
\mathrm{\gamma}_n(Q^2)\ {\cal M}_n(Q^2)\ ,
\label{eq:evolutionequationmoments}
\end{eqnarray}
which can be solved analytically.
Here the $n$-th full moment of the parton distribution is defined as:
\begin{eqnarray}
{\cal M}_n(Q^2) & = & 
\int_0^1 dx\; x^{n-1}\ q(x,Q^2)\ ,
\label{eq:fullmoment}
\end{eqnarray}
and the anomalous dimension,
\begin{eqnarray}
\mathrm{\gamma}_n\left(Q^{2}\right) & = & 
\int_0^1 dz\; z^{n-1} P\left(z,\alpha_S\left(Q^2\right)\right)\ ,
\label{eq:anomalousdimension}
\end{eqnarray}
is the moment of the splitting function
$P\left(z,\alpha_S\left(Q^2\right)\right)$.
The PDF can then be determined via the inverse Mellin transform,
\begin{eqnarray}
q(x,Q^2) & = & \frac{1}{2\pi i}
\int_{c-i\infty}^{c+i\infty}dn\; x^{-n}\
{\cal M}_n\left(Q^2\right)\ .
\label{eq:inversemellinfransform}
\end{eqnarray}

From the definition in Eq.~(\ref{eq:fullmoment}), the full moments are 
obtained by integrating the PDF over all values of the Bjorken variable, 
$0\leq x\leq1$.
Since $x$ is related to the invariant mass squared $W^{2}$ of the
virtual photon--hadron system, $W^2=M^2+Q^2\left(1-x\right)/x$,
where $M$ is the nucleon mass, to reach the $x \to 0$ limit requires
infinite energy; hence in practice some extrapolation to $x=0$ is
always needed to evaluate the moment.
Similarly, at finite $Q^2$ one usually excludes from leading twist
analyses the $W < 2$~GeV region in order to avoid low-$W$ nucleon
resonances, so an analogous extrapolation to $x=1$ is also performed.

\subsection{Truncated Moments}

An alternative approach, which avoids uncertainties from unmeasured
regions at low and high $x$, makes use of the so-called
``truncated moments'' \cite{Forte}.
In analogy with the full moments, the truncated moments of a PDF 
$q(x,Q^2)$ are defined as:
\begin{eqnarray}
{\cal M}_n(x_0,1,Q^2) & = & 
\int_{x_0}^1 dx\; x^{n-1}\ q(x,Q^2)\ ,
\label{eq:truncatedmoment}
\end{eqnarray}
where the integration is restricted to $x_0 \leq x \leq 1$
(the first two arguments in ${\cal M}_n$ denote the lower and upper
limits of the integration).
From the evolution equation (\ref{eq:evolutionequation}), one can
verify that the truncated moments satisfy:
\begin{eqnarray}
\frac{d{\cal M}_n(x_0,1,Q^2)}{dt} & = & 
\frac{\alpha_S\left(Q^2\right)}{2\pi}
\int_{x_0}^1 dy\; y^{n-1}\ q(y,Q^2)\
G_n\left(\frac{x_0}{y},Q^2\right)\ ,
\label{eq:evolutionequationtruncatedmoments2}
\end{eqnarray}
where
\begin{eqnarray}
G_n(x,Q^2) & = & 
\int_x^1dz\; z^{n-1} P\left(z,\alpha_S(Q^2)\right)
\label{eq:truncatedmomentofsplittingfunction}
\end{eqnarray}
is the truncated anomalous dimension.
For $x_0=0$, the latter reduces to the usual $x$-independent
anomalous dimension, $G_n(0,Q^2)=\gamma_n(Q^2)$,
which can be taken outside the integral in
Eq.~(\ref{eq:evolutionequationtruncatedmoments2}).
The right hand side then depends only on the $n$-th moment, and the
full moments of PDFs evolve independently of each other.

For nonzero $x_0$, the residual $y$ dependence in the truncated
anomalous dimension leads to evolutions equations which are not
diagonal in $n$.
This can be seen by expanding $G_n(x_0/y,Q^2)$ as a Taylor series 
around $y=1$, and truncating the expansion at a finite order $m$.
Accordingly, Eq.~(\ref{eq:evolutionequationtruncatedmoments2})
then turns into a system of coupled evolution equations:
\begin{eqnarray}
\frac{d{\cal M}_n(x_0,1,Q^2)}{dt} & = &
\frac{\alpha_S(Q^2)}{2\pi}
\sum_{k=0}^m c_{n,k}^{(m)}(x_0) {\cal M}_{n+k}(x_0,1,Q^2)\ ,
\label{eq:truncatedsum}
\end{eqnarray}
where
\begin{eqnarray}
c_{n,k}^{(m)}(x_0)=
\sum_{p=k}^m \frac{(-1)^{p+k} g_p^n(x_0)}{k!(p-k)!}
& 
\mathrm{and} & g_p^n(x_0) \equiv
\left.\frac{\partial^p}{\partial y^p}
	G_n\left(\frac{x_0}{y},Q^2\right)
\right|_{y=1}\ .
\label{eq:coefficientandderivative}
\end{eqnarray}

Unlike the full moments, the evolution of the truncated moment of order
$n$ is determined by all truncated moments of order $n+k$, with $k>0$.
However, the series of couplings to higher moments converges, and can
be truncated to any desired accuracy. 
One can solve Eq.~(\ref{eq:truncatedsum}) to arbitrarily high accuracy
by using a sufficiently large basis of truncated moments. 
For example, the higher moments $(n\geq2)$ can be calculated with
excellent accuracy even for a small $(m=4)$ number of terms in the
expansion of the truncated anomalous dimension.
The first moment, on the other hand, is more sensitive to the truncation
point $x_0$ and the convergence of the truncated anomalous dimension
for $n=1$ is weaker than for the higher moments.

\subsection{Diagonal Formulation of Truncated Moments}

The evolution equations satisfied by the truncated moments can be 
formulated in an alternative way which avoids the problem of mixing
of lower moments with higher moments \cite{Kotlorz}.
Inverting the order of integration on the right hand side of 
Eq.~(\ref{eq:evolutionequationtruncatedmoments2}) and introducing
a new variable $u = x_0 (y/x)$, the integral can be written as:
\begin{eqnarray}
\int_{x_0}^1 dx\; x^{n-1}\left(P\otimes q\right)(x,Q^2)
& = & \left(P'_n \otimes {\cal M}_n\right)(x_0,Q^2)\ ,
\label{eq:relationtruncatedmoments}
\end{eqnarray}
with
\begin{eqnarray}
P'_n(z,\alpha_S(Q^2))
& = & z^n P(z,\alpha_S(Q^2))\ .
\label{eq:pprime1}
\end{eqnarray}
The evolution equation for the truncated moments then becomes:
\begin{eqnarray}
\frac{d{\cal M}_n(x_0,1,Q^2)}{dt}
& = & \frac{\alpha_S(Q^2)}{2\pi}
\left( P'_n \otimes {\cal M}_n \right)(x_0,Q^2)\ ,
\label{eq:evolutionequationtruncatedmoments3}
\end{eqnarray}
which is very similar to the original evolution equation 
(\ref{eq:evolutionequation}) for the PDFs.
Here $P'_n$ plays the role of the splitting function for the
truncated moments.
The truncated moments therefore satisfy DGLAP evolution with a
modified splitting function 
$P(z,\alpha_S(Q^2)) \to z^n P(z,\alpha_S(Q^2))$
in the Mellin convolution.
The advantage of this approach is that it can be successfully applied
to any $n$-th moment and for every truncation point $0<x_0<1$,
without the complication of mixing with higher moments.

The evolution equations for the truncated moments can also be  
generalized to any subset in the $x$-region,
$x_{\rm min} \leq x \leq x_{\rm max}$.
Writing the ``doubly-truncated'' moment of the PDF as:
\begin{eqnarray}
{\cal M}_n(x_{\rm min},x_{\rm max},Q^2) & = & 
\int_{x_{\rm min}}^{x_{\rm max}} dx\; 
x^{n-1}\ q(x,Q^2)\ ,
\label{eq:doublytruncatedmoment}
\end{eqnarray}
its $Q^2$ evolution can be obtained by subtracting the solutions
of truncated moments at the points $x_{\rm min}$ and $x_{\rm max}$:
\begin{eqnarray}
{\cal M}_n\left(x_{\rm min},x_{\rm max},Q^2\right) & = &
{\cal M}_n\left(x_{\rm min},1,Q^2\right) -
{\cal M}_n\left(x_{\rm max},1,Q^2\right)\ ,
\label{eq:doublytruncatedmomentsolution}
\end{eqnarray}
where ${\cal M}_n(x_{\rm min},1,Q^2)$
and ${\cal M}_n(x_{\rm max},1,Q^2)$ both satisfy 
Eq.~(\ref{eq:evolutionequationtruncatedmoments3}).

\section{Data Analysis}
\label{sec:data}

The central aim of this study is to determine the extent to which
nucleon structure function data in specific regions in $x$ (or $W$)
are dominated by leading twist.
This can be done by constructing empirical truncated moments and
evolving them to a different $Q^2$ using one of two methods.
Namely, (i) the structure functions are evolved and the corresponding 
truncated moments the calculated, or (ii) the moments are evolved
directly using the evolution equations in 
Eq.~(\ref{eq:evolutionequationtruncatedmoments3}) above.
We found the results of both methods to be essentially equivalent.
In the study of the proton structure function data there is, however,
difficulty in applying the target mass corrections (TMCs) using the
latter method.
Here in principle one can derive and solve the evolution equations for 
the target mass corrected moments (the so-called truncated Nachtmann
moments), which contain the TMCs explicitly.
In practice, to avoid this problem we shall utilize method (i):
we evolve the structure functions, correct them for TMCs, and finally
calculate their moments.

Deviations of the evolved moments, computed to next-to-leading order
(NLO) accuracy, from the experimental data at the new $Q^2$ then
reveal any higher twist contributions in the original data. 
In particular, we will analyze recent data on the proton $F_2^p$ 
structure function from Jefferson Lab covering a range in $Q^2$ 
from $\lesssim 1$~GeV$^2$ to $\approx$~6~GeV$^2$.

The evolution of the measured truncated moments requires the structure
function to be decomposed into its nonsinglet and singlet components.
Without performing a global pQCD analysis of the structure function data,
it is {\em a priori} unknown which parts of the structure function are
singlet and nonsinglet.  To proceed, we shall assume that in our region 
of interest, at moderate to large $x$, the proton structure function 
is well approximated by its nonsinglet component, and will evolve the 
truncated moments as nonsinglets.  The accuracy of this approximation 
will improve with increasing order of the truncated moments.

\subsection{Evolution of Truncated Moments}

We test the accuracy of the nonsinglet evolution by first evolving a 
trial structure function whose decomposition into its nonsinglet and 
singlet components is known.
The trial function is evolved exactly, with its nonsinglet and singlet
components computed separately, and also evolved under the assumption
that the total function can be treated as a nonsinglet.
A comparison of the discrepancy between the two evolved truncated 
moments can then reveal the accuracy of the nonsinglet evolution of 
the various moments as a function of the truncation region.

There are many methods to solve the $Q^2$ evolution equations for the 
truncated moments \cite{Weigl}.
The simplest and most direct is to solve the equations by
brute force using a suitable numerical integration routine.
In this work we use the method of Ref.~\cite{Kumano} for the evolution.

To illustrate the method of direct moment evolution, we consider the
evolution of the nonsinglet truncated moment, ${\cal M}_n^{\rm NS}$,
to leading order in $\alpha_S$ (although in practice our numerical 
results are performed at NLO):
\begin{eqnarray}
\frac{d{\cal M}_n^{\rm NS}(x,1,\tau)}{d\tau} & = & 
\int_x^1 \frac{dy}{y}\;
\left(\frac{x}{y}\right)^n
P_{\rm NS}^{(0)}\left(\frac{x}{y}\right)
{\cal M}_n^{\rm NS}(y,1,\tau)\ ,
\label{eq:truncatedmomentxminleadingorder}
\end{eqnarray}
where the leading order NS splitting function is
\begin{eqnarray}
P_{\rm NS}^{(0)}(z) & = & 
\frac{4}{3}\left[\frac{1+z^2}{\left(1-z\right)_+} +
\frac{3}{2}\delta\left(1-z\right)\right]\ ,
\label{eq:nonsingletsplittingfunctionleadingorder}
\end{eqnarray}
and instead of $t$ we use the variable $\tau$, where
\begin{eqnarray}
\tau & \equiv & -\frac{2}{\beta_0}
\ln\left[\frac{\alpha_S(Q^2)}{\alpha_S(Q_0^2)}\right]\ ,
\label{eq:newvariabletau}
\end{eqnarray}
with $\alpha_S(Q^2) =
4\pi/\left[ \beta_0\ln(Q^2/\Lambda_{\rm QCD}^2) \right]$
the running coupling constant at leading order,
and $\beta_0 = 11 - 2N_f/3$ for $N_f$ quark flavors.
By dividing the variables $\tau$ and $x$ into small steps,
the evolution from $\tau_j$ to $\tau_{j+1}$ can be written as:
\begin{eqnarray}
{\cal M}_n^{\rm NS}(x_i,1,\tau_{j+1}) & = &
{\cal M}_n^{\rm NS}(x_i,1,\tau_j)	\nonumber\\
&+&
  \Delta\tau_j \sum_{k=i}^{N_x}
  \frac{\Delta x_k}{x_k} \left(\frac{x_i}{x_k}\right)^n
  P_{\rm NS}^{(0)}\left(\frac{x_i}{x_k}\right)
  {\cal M}_n^{\rm NS}(x_k,1,\tau_j)\ ,
\label{eq:numericalevolution}
\end{eqnarray}
where $\Delta\tau_j = \tau_{j+1} - \tau_j$ and
$\Delta x_k = x_k - x_{k-1}$ are the steps at positions $j$ and $k$, 
and $N_x$ is the number of steps in $x$.
The final truncated moment at $\tau_{N_{\tau}}$ is then obtained
by repeating the step in Eq.~(\ref{eq:numericalevolution})
$\left(N_{\tau}-1\right)$ times.

\begin{figure}[t]
\begin{center}
\includegraphics[scale=0.5]{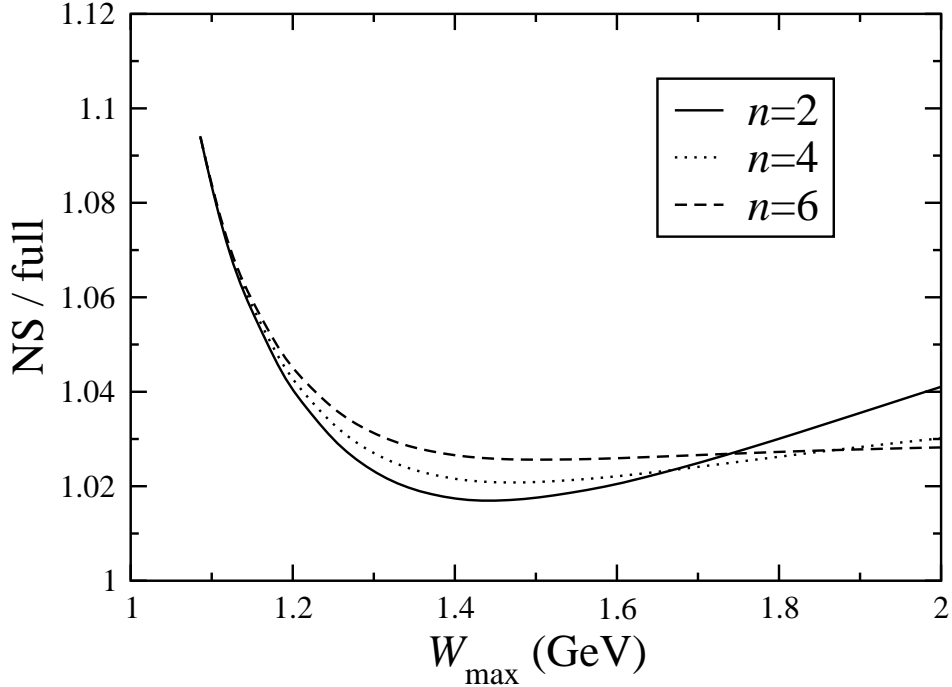}
\end{center}
\caption{Ratio of the truncated moments of $F_2^p$ evolved from
	$Q^2 = 25$ to 1~GeV$^2$, using NS and full evolution,
	versus the truncation point $W_{\rm max}$ (or $x_{\rm min}$),
	for the $n=2$ (solid), 4 (dotted) and 6 (dashed) moments.}
\label{fig:1}
\end{figure}

For the trial function we take the leading twist proton $F_2$
structure function computed from the MRST2004 PDF fit \cite{MRST2004}.
The $n=2$, 4 and 6 truncated moments of $F_2^p$ are then evolved from 
$Q^2 = 25$~GeV$^2$ to 1~GeV$^2$ using NS evolution, and compared with
the exact results using singlet and nonsinglet evolution.
The ratios of these are plotted in Fig.~\ref{fig:1}
as a function of the truncation point $W_{\rm max}$, where
$W_{\rm max}^2 = M^2 + Q^2 (1/x_{\rm min}-1)$.

Generally the differences between the full and NS evolution are of
the order 2--4\% for $1.2 \lesssim W_{\rm max} \lesssim 2$~GeV, the
traditional nucleon resonance region, and increase with increasing
$W_{\rm max}$.
Note that at $Q^2=1$~GeV$^2$, $W_{\rm max}=2$~GeV corresponds to 
$x_{\rm min}=0.24$.
For the $n=2$ moment, which is most sensitive to singlet evolution,
the differences do not exceed $\approx 4\%$ for
$1.2 \lesssim W_{\rm max} \lesssim 2$~GeV.
As expected, for the higher moments the differences are smaller,
$\lesssim 2-3\%$ for $n=4$ and $n=6$ for
$1.3 \lesssim W_{\rm max} \leq 2$~GeV.
In the region relevant for our study one can therefore safely conclude
that the error introduced by evolving the $F_2^p$ moments as nonsinglets
is less than 4\%.
This uncertainty will be included in the errors in our final results.

\begin{figure}[t]
\begin{center}
\includegraphics[scale=0.8]{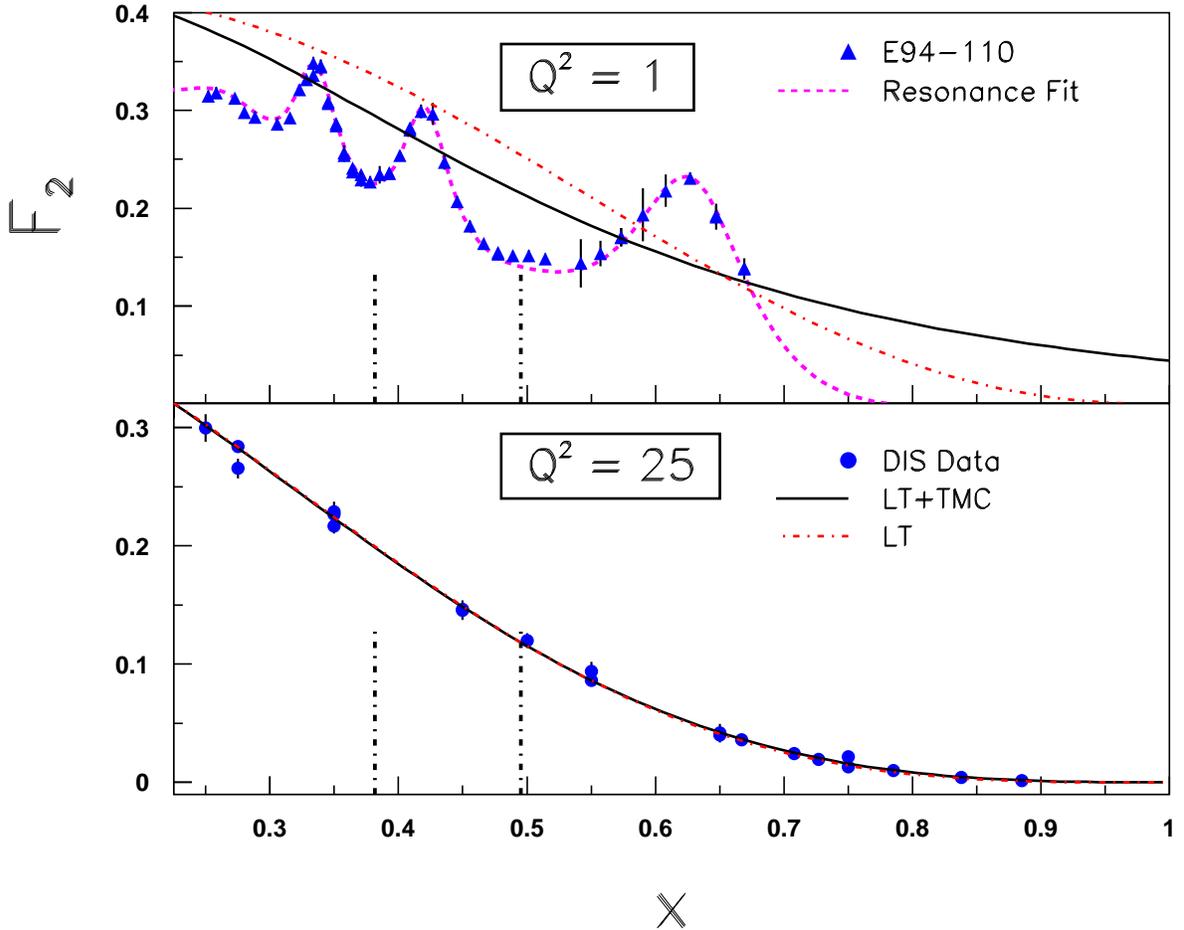}
\end{center}
\caption{{\em (Top panel)}
	Comparison of proton $F_2$ data from JLab experiment E91-110
	\cite{F2data} at $Q^2=1$~GeV$^2$ (triangles), and the fit
	\cite{Resfit} to the data (``Resonance Fit'', dashed),
	with a global fit of DIS data used to extract target mass
	contributions \cite{TMCfit}.
	Shown is the leading twist DIS fit with (``LT+TMC'', solid)
	and without (``LT'', dot-dashed) the TMCs.
	The vertical lines indicate the extent of the second ($S_{11}$)
	resonance region.
	{\em (Bottom panel)}
	DIS data at $Q^2=25$~GeV$^2$ (circles) compared with the LT
	and LT+TMC fits.}
\label{fig:2}
\end{figure}

\subsection{Extraction of Higher Twists}

Having tested the accuracy of the nonsinglet evolution we now turn
our attention to the analysis of the $F_2^p$ structure function data.
In Fig.~\ref{fig:2} (top panel) we compare the $F_2^p$ resonance data
from Jefferson Lab experiment E91-110 \cite{F2data} at $Q^2=1$~GeV$^2$
(triangles) with an empirical fit \cite{Resfit} to the data (dashed),
and with leading twist fits to the deep-inelastic $F_2^p$ data
\cite{TMCfit} with (solid) and without (dot-dashed) target mass
corrections.
The resonance fit \cite{Resfit} describes the $F_2^p$ data to better
than 3\% over the range $0 \leq Q^2 \leq 8$~GeV$^2$ and $W^2$ from
the inelastic threshold up to 10~GeV$^2$.

Since the data at low $Q^2$ contain significant contributions arising
from kinematical $M^2/Q^2$ corrections (which, although subleading in 
$Q^2$, contribute at leading twist), a direct comparison of data with
the leading twist structure function requires the inclusion of TMCs.
We do so here by applying the standard TMC prescription for $F_2$ from
Ref.~\cite{GP} (see also Ref.~\cite{TMC} for a review of TMCs).
As is evident from Fig.~\ref{fig:2}, the leading twist fit to the DIS
data, including TMCs, agrees well with the average $F_2^p$ data in each
resonance region.
A comparison of this fit with the DIS data at $Q^2=25$~GeV$^2$ (bottom
panel of Fig.~\ref{fig:2}) shows the excellent agreement between the 
leading twist function and the data at this scale.

More specifically, the comparison of the data with the target mass
corrected leading twist function illustrates the intriguing phenomenon
of Bloom-Gilman duality, where the data in the resonance region
oscillate around, and on average are approximately equal to, the
leading twist function \cite{BG}.
This duality reveals itself in the relatively small value of the higher
twist contributions at these scales observed in recent high-precision
$F_2^p$ measurements --- see Ref.~\cite{MEK} for a review of the data.
Note that the nonzero value of $F_2$ with TMCs in the limit $x \to 1$,
which is related to the so-called threshold problem, introduces a small
additional uncertainty into structure function analyses at low $Q^2$
\cite{TMC,Steffens}.

To determine the extent to which the $F_2^p$ data at low $Q^2$ are 
dominated by leading twist, we assume that the data beyond some large
$Q^2$ value are dominated by twist-2 contributions.
In view of the comparison with the data in Fig.~\ref{fig:2}, in this
analysis we take this scale to be $Q^2 = Q_0^2 = 25$~GeV$^2$.
This assumption is also consistent with most global analyses of PDFs,
which fit leading twist PDFs to structure function data down to
$Q^2 \sim 1$--2~GeV$^2$ \cite{MRST2004,CTEQ,Blumlein,NNPDF}.
Although these analyses typically exclude low-$W$ resonance data,
in practice there is little contribution to the low moments from
the resonance region $W \leq 2$~GeV.

The analysis method then proceeds in four main steps:
\begin{enumerate}
\item
For each $W^2$ region of interest, the $x$ range to be covered
($\Delta x$) is calculated at the particular (lower) $Q^2$ where the
leading twist contribution is to be extracted.
For the second ($S_{11}$) resonance region, for example, this is
indicated by the vertical lines in Fig.~\ref{fig:2} (top panel).
\item
The structure function extracted from a precision fit to data
\cite{TMCfit} at the starting scale $Q_0^2 = 25$~GeV$^2$ is evolved in
leading twist down to lower $Q^2$ and the TMCs are applied.
Note that at the higher $Q_0^2$ value the same interval $\Delta x$ is
used, as {\em e.g.} indicated by the vertical lines in Fig.~\ref{fig:2}
(bottom panel), which corresponds to a higher $W$ range.
\item
The truncated moment of the evolved leading twist, target mass corrected
structure function from Step~2 is calculated for the interval $\Delta x$
defined in Step~1.
\item
The truncated moment of the resonance data at $Q^2$ is calculated in the
interval $\Delta x$ and compared with the result of Step~3.
\end{enumerate}

\begin{figure}[t]
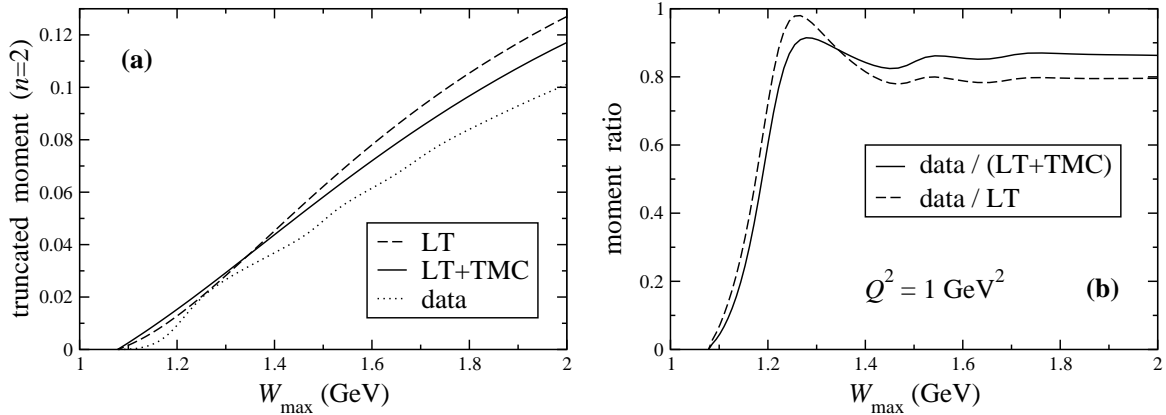

\begin{center}
\includegraphics[scale=0.3]{fig3a.eps}\ \ \
\includegraphics[scale=0.3]{fig3b.eps}
\end{center}
\caption{(a) Truncated moment ${\cal M}_2$ as a function of the
	truncation point $W_{\rm max}$ at $Q^2 = 1$~GeV$^2$,
	evolved as leading twist (LT) from $Q_0^2 = 25$~GeV$^2$ with
	(solid) and without (dashed) target mass corrections (TMC),
	and compared with the moment calculated from data (dotted) 
	at $Q^2 = 1$~GeV$^2$.
	(b) Ratio of the ${\cal M}_2$ truncated moments of the
	data to the leading twist + TMC (solid), and data to leading
	twist without TMC (dashed) at $Q^2 = 1$~GeV$^2$.}
\label{fig:3}
\end{figure}

After evolving down to $Q^2=1$~GeV$^2$ and applying the target mass
corrections, the $n=2$ truncated moment ${\cal M}_2$ is shown in 
Fig.~\ref{fig:3}(a) as a function of $W_{\rm max}$, where it is
compared with the moment of the actual data at $Q^2=1$~GeV$^2$.
The difference between the evolved curve and the data attests to the 
presence of higher twist contributions in the data at $Q^2=1$~GeV$^2$.
The importance of the TMCs is also clearly evident, and these in fact
reduce the difference between the leading twist moment and the data
by some 40\% for large $W_{\rm max}$.
The ratio of the truncated moments of the data to the leading twist in
Fig.~\ref{fig:3}(b) illustrates that without TMCs the leading twist
moment differs from the data by $\sim 20\%$ for $W_{\rm max} > 1.5$~GeV.
After correcting for TMCs, the size of the apparent higher twists is
reduced to $\sim 15\%$.
It is imperative, therefore, that the kinematical effects associated
with finite values of $Q^2/\nu^2$ be properly accounted for before
drawing any conclusions about higher twists from data.

Note that the truncated moments displayed in Fig.~\ref{fig:3} are
computed over the range $W_{\rm th} \leq W \leq W_{\rm max}$, where
the $W_{\rm th} = M + m_\pi$ is the inelastic threshold.
This is consistent with the assumption that the truncated moments at
$Q_0^2 = 25$~GeV$^2$ are entirely of twist-two, since the elastic cross 
section contributes only to the higher twist part of the structure
function.
For a meaningful comparison, we therefore do not include the elastic
contribution at lower $Q^2$.
At $Q^2 = 1$~GeV$^2$ this corresponds to the integration range
$x_{\rm min} \leq x \leq x_{\rm th}$, where 
$x_{\rm th} = \left[ 1 + m_\pi (m_\pi+2M)/Q^2\right]^{-1} \simeq 0.78$.

\begin{figure}[t]
\begin{center}
\includegraphics[scale=0.55]{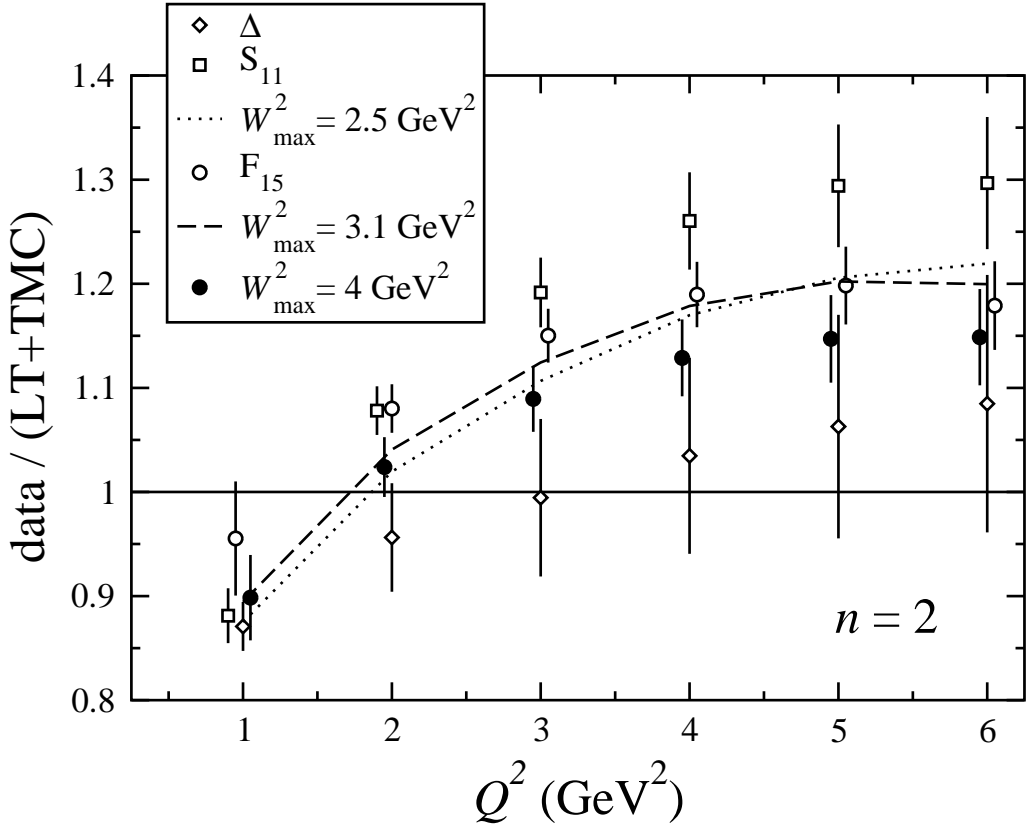}
\end{center}
\caption{$Q^2$ dependence of the ratio of $n=2$ truncated moments
	${\cal M}_2$ calculated from the data and from leading twist
	evolution from $Q_0^2 = 25$~GeV$^2$ (including TMCs), for
	various intervals in $W$: the first ($\Delta$) resonance region
	(diamonds), second ($S_{11}$) resonance region (squares),
	the first and second combined, corresponding to
	$W_{\rm max}^2 = 2.5$~GeV$^2$ (dotted curve), third ($F_{15}$)
	resonance region (open circles), first three regions combined,
	$W_{\rm max}^2 = 3.1$~GeV$^2$ (dashed curve), and the entire
	resonance region $W_{\rm max}^2 = 4$~GeV$^2$ (filled circles).
	Note that some of the points are offset slightly for clarity.}
\label{fig:4}
\end{figure}

\begin{figure}[t]
\begin{center}
\includegraphics[scale=0.55]{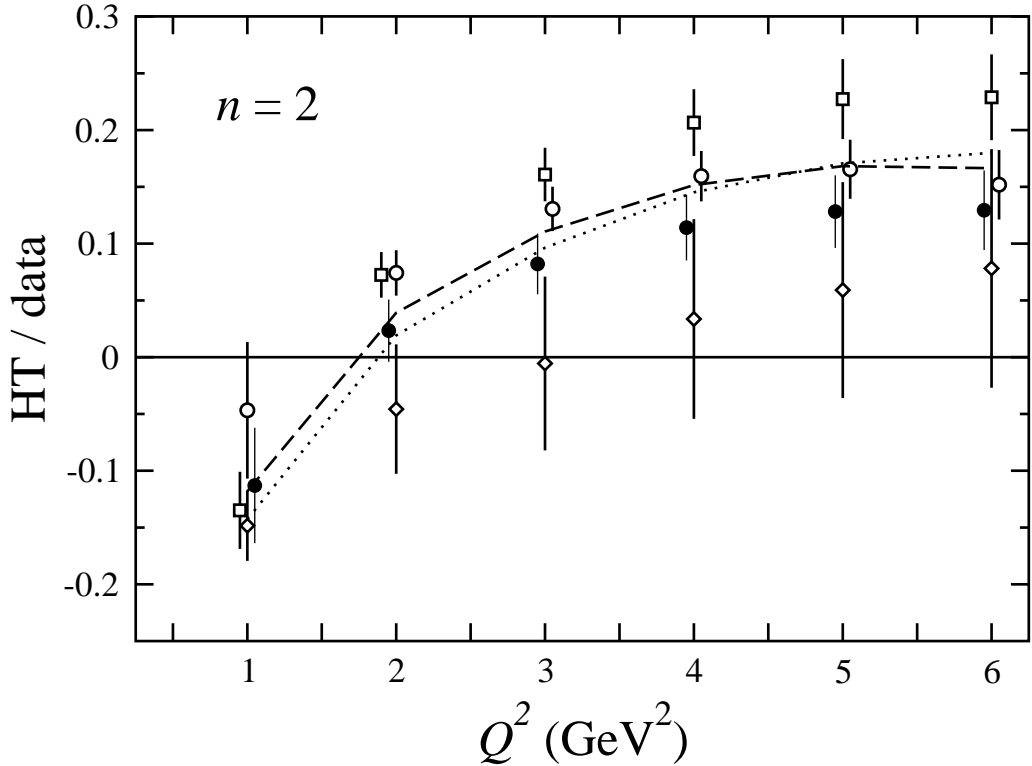}
\end{center}
\caption{$Q^2$ dependence of the fractional higher twist (HT) 
	contribution to the $n=2$ truncated moment data, for various 
	intervals in $W$ (as in Fig.~\ref{fig:4}).}
\label{fig:5}
\end{figure}

The results in Fig.~\ref{fig:3} give a clear indication of the magnitude 
and sign of higher twists in the data at $Q^2 = 1$~GeV$^2$.
To quantify the higher twist content of the specific resonance regions,
and at different values of $Q^2$, we consider several intervals in $W$:
$W_{\rm th}^2 \leq W^2 \leq$~1.9~GeV$^2$, corresponding to
the traditional $\Delta(1232)$ (or first) resonance region;
$1.9 \leq W^2 \leq 2.5$~GeV$^2$ for the $S_{11}(1535)$ (or second)
resonance region; and
$2.5 \leq W^2 \leq 3.1$~GeV$^2$ for the $F_{15}(1680)$ (or third)
resonance region.
The $n=2$ truncated moments corresponding to these regions are plotted
in Fig.~\ref{fig:4} for various $Q^2$ values, from $Q^2 = 1$~GeV$^2$ 
to $Q^2 = 6$~GeV$^2$.
Shown are ratios of moments calculated from the data to the moments
obtained from NLO evolution of the leading twist moments from
$Q_0^2 = 25$~GeV$^2$, corrected for target mass effects.
Below $Q^2 = 1$~GeV$^2$ the applicability of a pQCD analysis becomes 
doubtful and the decomposition into leading and higher twists is no 
longer reliable.

The results indicate deviations from leading twist behavior of the
entire resonance region data (filled circles in Fig.~\ref{fig:4})
at the level of $\lesssim 15\%$ for all values of $Q^2$ considered,
with significant $Q^2$ dependence for $Q^2 \lesssim 4$~GeV$^2$.
This is made more explicit in Fig.~\ref{fig:5}, where the higher twist 
contributions to ${\cal M}_2$ (defined as the difference between the 
total and leading twist moments) are shown as ratios of the moments 
evaluated from the data.

The strong $Q^2$ dependence of the higher twists is evident here in
the change of sign around $Q^2 = 2$~GeV$^2$, with the higher twists
going from $\approx -10\%$ at $Q^2 = 1$~GeV$^2$ to $\approx 10$--15\%
for $Q^2 \approx 5$~GeV$^2$.
The slope at $Q^2 \approx 1-2$~GeV$^2$ would be decreased if the full
NS + singlet evolution were performed, as evident from Fig.~\ref{fig:1},
since the NS-only evolution leads to a few percent overestimate of the
LT+TMC results.
At larger $Q^2$ the higher twists are naturally expected to decrease,
once the leading twist component of the moments begins to dominate.
Note that the extraction of higher twists beyond $Q^2 = 6$~GeV$^2$
would require evolution from a starting scale larger than the
$Q_0^2 = 25$~GeV$^2$ used in this analysis.
At larger $Q^2$, however, data in the large-$x$ region, which
determines the behavior of the resonances after evolution to lower
$Q^2$, are not well determined, making extraction of higher twists
beyond $Q^2 \approx 6$~GeV$^2$ problematic at present.

Turning to the individual resonance regions, the results in
Figs.~\ref{fig:4} and \ref{fig:5} show that in the $\Delta$ region
(diamonds) the higher twist contributions are smallest in magnitude
at large $Q^2$, decreasing from $\approx -15\%$ of the data at 
$Q^2=1$~GeV$^2$ to values consistent with zero (within errors) at 
larger $Q^2$.
The higher twists are largest, on the other hand, for the $S_{11}$
region (squares), where they vary between $\approx -15\%$ of the
data at $Q^2=1$~GeV$^2$ and 20--25\% at $Q^2 \sim 5$~GeV$^2$.
Combined, the higher twist contribution from the first two resonance
regions (dotted curve) is $\lesssim 15\%$ in magnitude for all $Q^2$.
The rather dramatic difference between the $\Delta$ and the $S_{11}$,
may, at least in part, be due to the choice of the differentiation
point of $W^2 = 1.9$~GeV$^2$.
A lower $W^2$ choice, for instance, would lower the higher twist
content of the $S_{11}$ at large $Q^2$, while raising that of the
$\Delta$.
However, our $W^2$ choice corresponds to the local minimum between
these two resonances in the inclusive spectra, and is the one most
widely utilized.

The higher twist content of the $F_{15}$ region (open circles) is
similar to the $S_{11}$ at low $Q^2$, but decreases more rapidly
for $Q^2 > 3$~GeV$^2$.
The higher twist content of the first three resonance regions
combined (dashed curve) is $\lesssim 15$--20\% in magnitude for
$Q^2 \leq 6$~GeV$^2$.
Integrating up to $W_{\rm max}^2 = 4$~GeV$^2$ (filled circles),
the data on the $n=2$ truncated moment are found to be leading twist
dominated at the level of 85--90\% over the entire $Q^2$ range.

The results in Figs.~\ref{fig:4} and \ref{fig:5} contain two sources
of uncertainty: from the experimental uncertainty on the $F_2$ data 
(statistical and systematic), and from the nonsinglet evolution of the 
data.
For the experimental error we take an overall uncertainty of 2\% for
all truncated moment data, with the exception of the $n=4$ and $n=6$
moments for $W_{\rm max}^2 = 1.9$ and 4~GeV$^2$, where the experimental 
uncertainties are 4\% and 3\% for ${\cal M}_4$, and 5\% and 4.5\% for 
${\cal M}_6$ for the two $W$ regions, respectively.
The evolution error is estimated by comparing the nonsinglet evolution 
with the full evolution using the MRST fit, as in Fig.~\ref{fig:1},
with the appropriate correction factor applied at each $Q^2$ and $W$
interval.
We do not assign an uncertainty for the structure function at the
$Q_0^2 = 25$~GeV$^2$ input scale, as this is negligible for all but
the $\Delta$ region analysis at $Q^2 = 6$~GeV$^2$, where we estimate
it to be $\lesssim 3\%$.

Note that the size of the error bars on the $S_{11}$ and $F_{15}$ data
points at the lowest $Q^2$ are larger than those for the $\Delta$.
This is due to the fact that these resonances lie at higher $W$, and
hence at lower values of $x$ compared to the $\Delta$, and where the 
NS approximation to the full evolution is not as good as at large $x$.
The error bar on the $\Delta$ is smaller at the same low $Q^2$ values
since it appears at larger $x$, where the NS versus full evolution
differences are smaller.

The relatively small size of the higher twists at scales
$\sim$~few~GeV$^2$ is consistent with the qualitative observations
made in earlier data analyses about the approximate validity of 
Bloom-Gilman duality \cite{Niculescu}.
In this analysis we are able to for the first time quantify precisely
the degree to which this duality holds as a function of $Q^2$
(see also Ref.~\cite{Liuti}).
The fact that duality works better (higher twists are smaller) when
more resonances are included is also borne out in various quark model
studies \cite{Quarks,MEK}.

\begin{figure}[t]
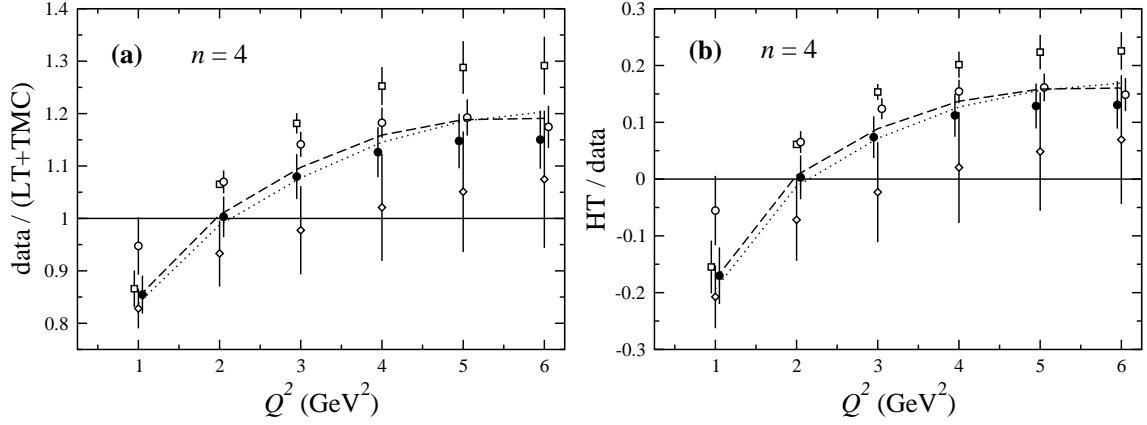

\begin{center}
\includegraphics[scale=0.3]{fig6a.eps}\ \
\includegraphics[scale=0.3]{fig6b.eps}
\end{center}
\caption{(a) $Q^2$ dependence of the ratio of truncated moments
        ${\cal M}_4$ calculated from the data and from leading twist
        evolution from $Q_0^2 = 25$~GeV$^2$ (including TMCs), for
	various intervals in $W$ (labels as in Fig.~\ref{fig:4}).
	(b) Fractional higher twist contribution to the $n=4$
	truncated moment data, for various intervals in $W$
        (as in Fig.~\ref{fig:5}). \\ }
\label{fig:6}
\end{figure}

\begin{figure}[t]
\begin{center}
\includegraphics[scale=0.3]{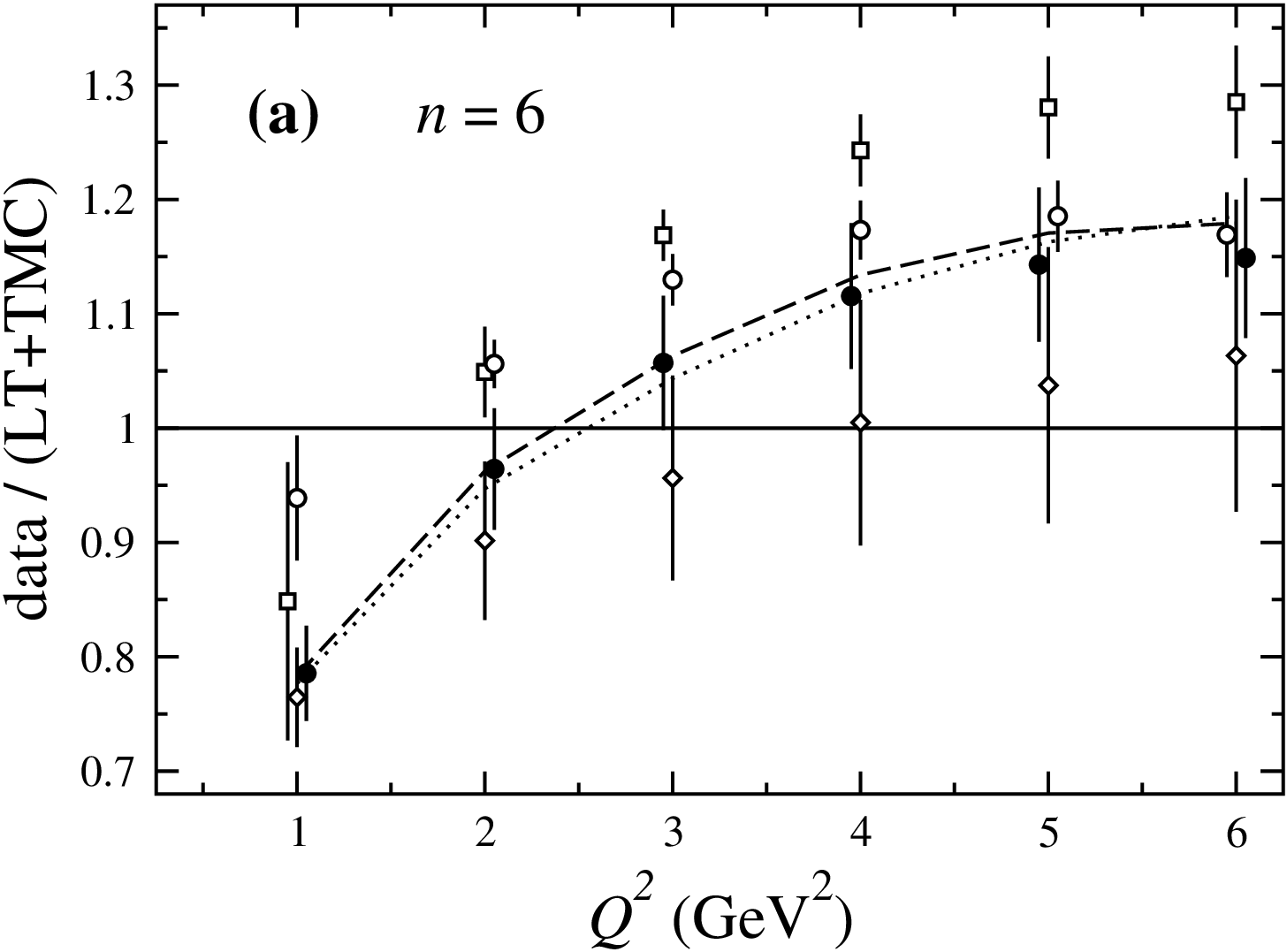}\ \
\includegraphics[scale=0.3]{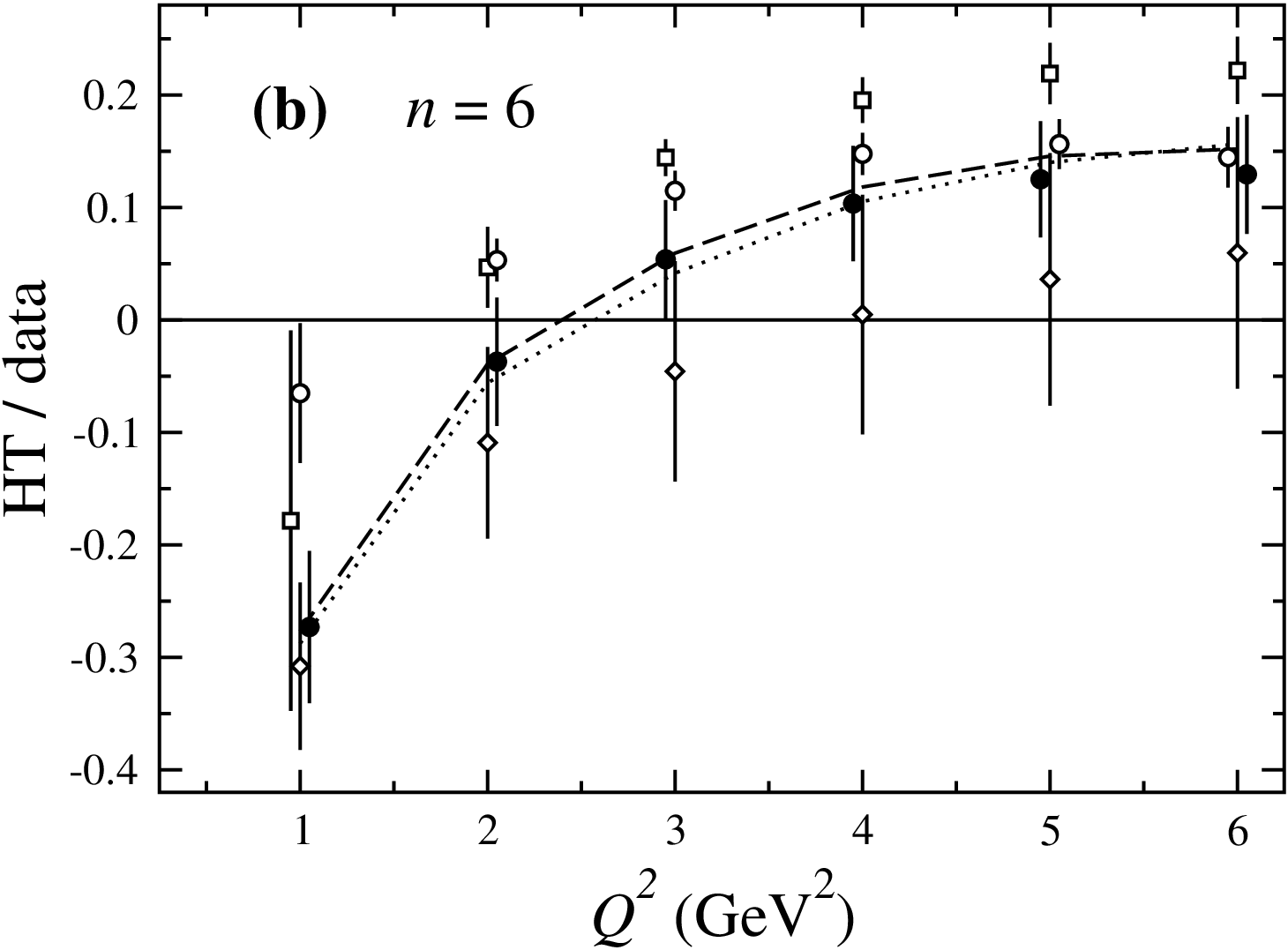}
\end{center}
\caption{(a) Ratio of truncated moments ${\cal M}_6$ calculated from the
	data and from leading twist evolution from $Q_0^2 = 25$~GeV$^2$ 
	(including TMCs), for various intervals in $W$ (labels as in 
	Fig.~\ref{fig:4}).
	(b) Fractional higher twist contribution to the $n=6$
	truncated moment data, for various intervals in $W$
	(as in Fig.~\ref{fig:5}).}
\label{fig:7}
\end{figure}

Similar behavior is found also for the $n=4$ and $n=6$ truncated moment 
ratios, illustrated in Figs.~\ref{fig:6} and \ref{fig:7}, respectively.
For the higher moments, the overall magnitude of the higher twists is
qualitatively similar to the $n=2$ moments, although the $Q^2$ values at
which they start decreasing in importance are larger.
At low $Q^2$ values the higher twist contributions are also relatively
larger for higher moments: at $Q^2 = 1$~GeV$^2$, for example, the
magnitude of the higher twist component of the $W^2 < 4$~GeV$^2$ region
increases from $\sim 10\%$ for the $n=2$ moment, to $\sim 15$--20\% for
$n=4$, and $\sim 20$--30\% for $n=6$.
This behavior can be understood from the relatively greater role played
by the nucleon resonances and the large-$x$ region, which is emphasized
more by the ($x$-weighted) higher moments.

\section{Conclusion}
\label{sec:conclusion}

Quark-hadron duality in nucleon structure functions remains an
intriguing empirical phenomenon which challenges our understanding of 
strong interaction dynamics, as one transitions from hadronic degrees 
of freedom in the nucleon resonance region to quarks and gluons in 
the deep inelastic continuum.
Until now the only rigorous connection with QCD has been for moments
of structure functions analyzed within the twist expansion.
Any insight about the workings of duality for individual resonances,
or specific resonance regions, has been confined to QCD-inspired
models of the nucleon.

In this paper we have presented a new approach to the study of local 
quark-hadron duality within a perturbative QCD context, using so-called
{\em truncated} moments of structure functions.
The fact that truncated moments obey a similar set of $Q^2$ evolution 
equations to the DGLAP equations for parton distributions enables a 
rigorous connection to be made between local quark-hadron duality and 
QCD.
It allows us to quantify for the first time the higher twist content
of various resonance regions, and determine the degree to which 
individual resonance regions are dominated by leading twist.

We find deviations from leading twist behavior of the truncated
moments of the resonance region data ($W \leq 2$~GeV) at the level
of $\lesssim 15\%$ for $Q^2 > 1$~GeV$^2$.
Significant $Q^2$ dependence in the ratio of moments of data to
leading twist is evident for $Q^2 \lesssim 3$~GeV$^2$, with the
higher twists changing sign around $Q^2 = 2$~GeV$^2$.
For the $n=2$ truncated moment, ${\cal M}_2$, the higher twists
are found to vary from $\approx -10\%$ at $Q^2 = 1$~GeV$^2$ to
$\approx 10$--15\% at $Q^2 \approx 5$~GeV$^2$.

Separating the $W \leq 2$~GeV data into the three traditional
resonance regions, our results indicate that at a scale of
$Q^2 = 1$~GeV$^2$ the $\Delta$ resonance region contains about
$-15\%$ higher twist component of the total ${\cal M}_2$, but
is consistent with zero at larger $Q^2$.
The higher twists in the second $(S_{11})$ and third $(F_{15})$
resonance regions are larger in magnitude, with the $S_{11}$
ranging from $\approx -15\%$ at $Q^2=1$~GeV$^2$ to 20--25\% at
$Q^2 \sim 5$~GeV$^2$, and the $F_{15}$ varying from 0 and 15\%
over the same range.

Similar behavior is found also for the $n=4$ and $n=6$ truncated moments.
Here the relatively greater role played by the resonances due to the
large-$x$ enhancement leads to larger higher twists at the same $Q^2$.
At $Q^2 = 1$~GeV$^2$, for example, the higher twist component of the
$W \leq 2$~GeV region increases from around $-10\%$ for ${\cal M}_2$ to
$\approx -15\%$ for ${\cal M}_4$, and $\approx -25\%$ for ${\cal M}_6$.

In contrast to earlier analyses of duality using complete moments of
structure functions which have quantified the total higher twist content
over all $x$, this analysis in terms of truncated moments reveals the
distribution of higher twist corrections over various regions in $x$
(or $W$).
Note that, unlike many previous moment analyses, an effort was also
made to quantify the uncertainty associated with evolving the structure
function data as a nonsinglet, which was found to be $\lesssim 4\%$.

While this analysis has been to some extent exploratory, it has
illustrated an encouraging new approach to quantifying and
understanding local Bloom-Gilman duality within a well-defined
theoretical framework.
It opens the way to further study of local duality in other structure
functions, such as the longitudinal structure function $F_L$ or
spin-dependent structure functions.

\begin{acknowledgments}
We thank I.~Clo\"et for helpful communications, and S.~Kumano for
providing the $Q^2$ evolution code from Ref.~\cite{Kumano}.
This work was supported by the U. S. Department of Energy contract
DE-AC05-06OR23177, under which Jefferson Science Associates, LLC operates
the Thomas Jefferson National Accelerator Facility, and National Science
Foundation grant 0400332.
\end{acknowledgments}


\end{document}